\documentclass[preprint]{elsarticle}
\usepackage{lineno,hyperref}
\usepackage{xcolor}
\usepackage{subcaption}
\usepackage{graphicx} 
\usepackage{siunitx}
\usepackage{amsmath}
\usepackage{booktabs}
\usepackage{dirtytalk}
\usepackage{enumitem}
\RequirePackage{snapshot}
\graphicspath{ {./images/} }

\hypersetup{
  colorlinks=true,
}

\makeatletter
\def\ps@pprintTitle{%
  \let\@oddhead\@empty
  \let\@evenhead\@empty
  \def\@oddfoot{\reset@font\hfil\thepage\hfil}
  \let\@evenfoot\@oddfoot
}
\makeatother

\begin{document} 

\hypersetup{
  linkcolor=blue,
  urlcolor=blue,
  citecolor=blue
}

\begin{frontmatter}

\title{Affordable inline structuration measurements of printable mortar with a pocket shear vane} 

\author[urn]{L\'eo Demont} 
\author[enpc,urn]{Romain Mesnil} 
\author[XtreeE]{Nicolas Ducoulombier} 
\author[urn]{Jean-Fran\c cois Caron \corref{cor1}}

\address[urn]{Navier Laboratory, Ecole des Ponts ParisTech, Univ. Gustave Eiffel, CNRS, Marne-La-Vall\'ee, France} 
\address[enpc]{Build'In, \'Ecole des Ponts ParisTech, Marne-La-Vall\'ee, France} 
\address[XtreeE]{XtreeE, 18 Rue du Jura, 94150 Rungis}
\cortext[cor1]{Corresponding author\\ Email: \href{mailto:jean-francois.caron@enpc.fr}{jean-francois.caron@enpc.fr}} 

\begin{abstract}
The control of mortar rheology is of paramount importance in the design of systems and structures in 3D printing concrete by extrusion. This is particularly sensitive for two-component 
 (2K) processes that use an accelerator to switch the printed mortar very quickly from a liquid behavior to a sufficiently solid behavior to be able to be printed. It is necessary to set up simple and effective tests within a precise methodological framework to qualify materials evolving so quickly in an industrial context. It is obvious that inline solutions, that is to say, post-printing solutions, will be more desirable than benchtop-type solutions reproducing the printing conditions as well as possible, but imperfectly. After some main key points about measuring the structuration of mortars, we propose an original inline test using a pocket shear vane tester. The protocols are precisely described and the simplicity and quality of the results are demonstrated.
\end{abstract}

\begin{keyword}
\texttt additive manufacturing \sep  concrete 3D printing \sep pocket shear vane \sep yield stress  \sep build-up rate \sep inline measurement
\end{keyword}

\end{frontmatter}

\section{Introduction}

The control of mortar rheology is of paramount importance in the design of systems and structures in 3D printing concrete by extrusion. The main rheological requirements for a successful implementation of the process are detailed in a seminal paper by Roussel \cite{roussel_rheological_2018}, where the importance of yield stress evolution is illustrated with the analytical solution of a vertical wall. The \textit{buildability requirement} states that the stress resulting from self-weight $\rho g$ does not exceed the yield stress $\tau_c$ of the material, as shown in equation \eqref{eq:wall} and in Figure \ref{fig:buildability_bicomponent}, where it is assumed that the total height $H(t)$ is a linear function of time.
\begin{equation}
    \tau_{cr}\left(t\right)>\frac{\rho g H\left(t\right)}{\sqrt{3}}
    \label{eq:wall}
\end{equation}
If the build-up rate (or structuration) is not sufficient, the limit of constructability is reached (red zone in Figure \ref{fig:buildability_bicomponent}) and printing is no longer possible because the first layers cannot support the load of the successive ones.
Moreover, a rapid build-up rate (or structuration) is then required to increase productivity. For example, one gets $A_{thix} \simeq 25$ Pa/s (from  Eq. \eqref{eq:structuration_vitesse}) for a vertical speed $V_r=6m/hour$. As a result, the yield stress of a printable material spans several decades during the first hours of printing \cite{roussel_printable_2022}. The build-up rate is also important in the context of the fiber-reinforced printed mortars, the impregnation of the fibers will be better ensured by a fluid material, which then quickly structures itself (see for example the process \textit{Flow-Based Pultrusion} \cite{CARON_2021}\cite{ducoulombier_additive_2020} \cite{demont_flow-based_2021} for more details). More recently, Carneau and coauthors proposed a stability criterion based on the structuration rate for the layer-pressing problem and showed that local layer geometry may be impacted by its value \cite{carneau_layer_2022}.

\subsection{Rheological behavior of printable mortars}
As a general rule, the evolution of mortar or any traditional cementitious paste rheological properties can be separated into two stages: the dormant period which generally corresponds to a printable mortar to the first hour after it has been mixed, which we will call the very young age, then the setting time characterized by an exponential acceleration of the hydration reactions and therefore of the structuration, as shown in \cite{perrot_structural_2016-1}. 

For the dormant period, Roussel proposed a linear evolution of the yield stress (Eq. \ref{eq:athix}) in \cite{roussel_thixotropy_2006-1}, which captures well the very young age behavior. 
\begin{equation}
    \tau_{cr}(t) = \tau_0 + t \cdot A_{thix}
    \label{eq:athix}
\end{equation}

Perrot \textit{et al.} proposed a nonlinear model for the time evolution of yield stress, with the introduction of a characteristic time $t_c$ \cite{perrot2015prediction}, shown in equation \eqref{eq:structuration_perrot}. The merit of this model is to converge towards an exponential model for a large time, thus capturing hydration, but it also allows fixing the slope at $\left(t=0\right)$.   
\begin{equation}
    \tau_{cr}\left(t\right)=A_{thix}t_c\left(e^{t/t_c}-1\right)+\tau_0
    \label{eq:structuration_perrot}
\end{equation}

Note that these equations hold only after the re-flocculation phase that occurs just after the material is set to rest, and that other researchers observe a re-flocculation rate much higher than the structuration rate \cite{kruger_ab_2019}.

Assuming a linear evolution of the yield stress over time, the build-up rate requirement can thus be written as follows, assuming a constant vertical speed $V_z$ and vertical printed walls.
\begin{equation}
    \tau_0+\left(A_{thix}-\frac{\rho g V_z}{\sqrt{3}}\right)t>0
    \label{eq:structuration_vitesse}
\end{equation}
If one neglects the initial yield stress, which is usually low in the context of bi-component (2K) concrete 3D printing, the buildability requirements just write $\left(A_{thix}-\frac{\rho g V_z}{\sqrt{3}}\right)>0$. In other words, the maximal vertical speed is directly proportional to the structuration rate $A_{thix}$. 

\begin{figure}[htbp] 
    \centering 
   \includegraphics[width=1\columnwidth]{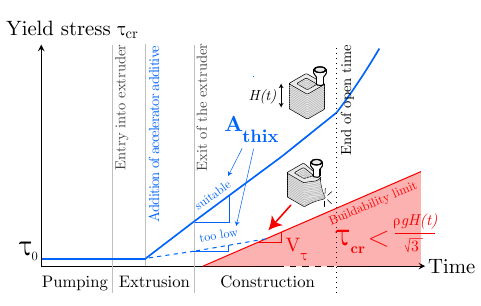} 
    \caption{Evolution of shear threshold $\tau_{cr}$ of a 2K printed material.} 
    \label{fig:buildability_bicomponent} 
\end{figure}

In the bi-component (2K) technology, such as the one developed by XtreeE and used in the Build'in platform, the mortar has a low yield stress that is nearly constant over time ($A_{thix}$ is small) to efficiently handle the pumping stage, even for high plastic viscosity mortar such as UHPC. Then, the increase of rheological properties is started ($A_{thix}$ increasing) through the addition of an accelerator using a dosing and mixing device located in the printing head just before extrusion \cite{gosselin_large-scale_2016-1,esnault_online_2017,esnault_experience_2019}. The schematic evolution of rheological properties over time of such a process is illustrated in Figure \ref{fig:buildability_bicomponent}. With the use of an appropriate additive dosing, the material timeline (continuous blue curve) has a suitable build-up rate $A_{thix}$ that allows respecting the buildability limit (Eq. \ref{eq:wall}) shown in red. On the contrary, insufficient dosing means an insufficient build-up: the material timeline (dashed blue curve) goes below the buildability limit and fails by plastic flow as shown in the pictogram. Different accelerators for 3D printing concrete exist in the literature, but in this work, an alkali-free aluminum sulfate solution is used. From a physicochemical point of view, the addition of alkali-free additives triggers the rapid formation of ettringite needles, due to the reaction between aluminum sulfate and calcium \cite{bravo2003effects}. Those needles jam the internal mortar microstructure which results in an increase in yield stress over time. 

Dressler and coauthors showed that the accelerator concentration had a significant impact on the early-age structuration of printable mortar \cite{dressler_effect_2020}. However, the effect of the accelerator dosage on the build-up rate is still an active topic of research, as the description of the interaction between plasticizer, accelerator, cement, and the compactness of the granular skeleton \cite{LOWKE201894} needs to be clarified. However, note that this type of accelerator does not necessarily reduce the dormant period, after which occurs the classic hydration phenomena leading to the acceleration of the CSH formation, which usually happens seven hours after mixing with water in our case. The development of systematic and representative inline characterization methods of the printed material is thus necessary.
In this paper, we will then restrict our analysis to this linear increase of yield stress through time and assume that the addition of accelerator is only increasing the slope of the linear increase of the yield stress over time $A_{thix}$, which simplifies the analytical derivation.\\

\subsection{Measuring build-up rate}
Several methods for measuring the build-up rate have been proposed for printing mortars \cite{esnault_experience_2019}, but the question arises whether they are applicable in the form of systematic control tests at a very young age representative of the true two-component printing process. \\ 

\begin{figure}[htbp] 
    \centering
    \includegraphics[width=1\columnwidth]{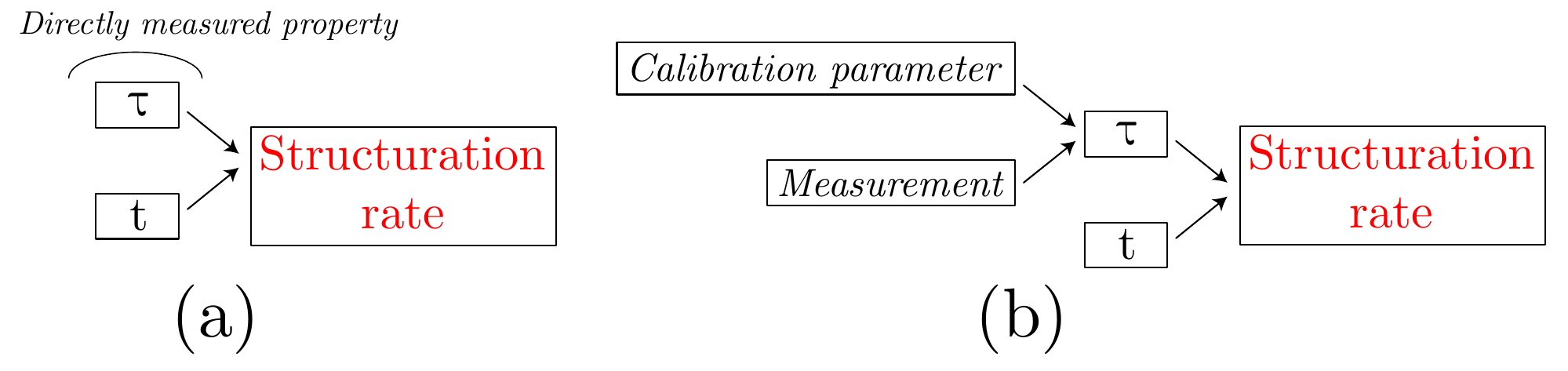} 
    \caption{Diagram of the principles of direct (a) and indirect (b) measurement of the structuration rate.} 
    \label{fig:direct_vs_indirect} 
\end{figure}

The fast evolution of yield stress makes characterization difficult. Promising methods that can measure the yield stress of printable materials can be found in \cite{roussel_printable_2022,reiter_role_2018}. We will focus here on so-called direct or destructive techniques (Figure \ref{fig:direct_vs_indirect}a). With these, the time evolution of the yield stress $\tau_{cr}$ associated with rupture or the onset of flow, is obtained via one or several measurements. We, therefore, do not discuss indirect techniques, such as ultrasound measurements, where the threshold $\tau_{cr}$ is calculated as a function of a material property $p$ and rheological data $c$ specific to the material and obtained by another direct test \cite{wolfs_correlation_2018} (Figure \ref{fig:direct_vs_indirect}b), nor non-destructive mechanical measurements, such as static Vicat plate or needle tests \cite{amziane_novel_2008}\cite{sleiman_new_2010} which do not directly measure the yield stress (or rupture) related to the collapse of the internal granular system. 
In the next section, we review and precise some specific and important aspects which qualify a given test protocol. After that, we will introduce and detail the pocket hand vane test proposal.  

\section{Principal methods and some key points for structuration rate measurements.}

\label{section:bench}
 We highlight here some important points for the qualification of a test aiming to measure the build-up rate of a mortar. Some definitions are given to help define the limitations of a given test. A rheological test can be characterized by several steps:
\begin{itemize}
    \item A sample of volume $\Omega$ is prepared, ideally by using a material that has been pumped by the extrusion process, in a time $t_{sample}$
    \item After a sufficient time $t_{rest}$, for dispersing the accelerators and for the structuration to be initiated at the moment of the measurement  \cite{roussel_thixotropy_2006-1}\cite{lootens_ciments_nodate}, the test phase can begin, and $\tau_{0}$ is the initial threshold at this instant.
    \item The test is prepared, for example by putting the material sample on a testing machine, and performed, which requires at least the time $t_{min}$
\end{itemize}
If one considers inline process, the time $t_{sample}$ necessary for preparation is simply equal to $\Omega/Q$, where $Q$ is the flow-rate of the extruder: it is thus system dependent and is not an intrinsic property of the test. The flow rate may vary significantly depending on the process. A good bulk approximation sets $Q$ between $1-10 L/min$ for many 3d printing processes. In addition, every test can measure the yield stress between two bounds $[\tau_{min},\tau_{max}]$.

Table \ref{tab:test map} displays the characteristics for some popular tests listed in \cite{nicolas_assessing_2022-1}. Some tests are \textit{gravity-driven}, like the slug-test \cite{ducoulombier_slugs-test_2021}, Abrams cone, or mini-cone. Other are \textit{gravity-dependent}, like the compression test. It shall be noted that the slug-test is by design a test with no rest time. It is also the only test where the notion of preparation time $t_{sample}$ is not relevant since the material tested is always the material at the nozzle exit. 

\begin{table}[ht!]
\begin{tabular}{@{}lccccccc@{}}
\toprule
\multicolumn{1}{c}{\textbf{Test}} & \textbf{\begin{tabular}[c]{@{}c@{}}Gravity\\ dependent\end{tabular}} & \textbf{$\tau_{min}$} & \textbf{$\tau_{max}$} & \textbf{\begin{tabular}[c]{@{}c@{}}H\\ {[}mm{]}\end{tabular}} & \textbf{\begin{tabular}[c]{@{}c@{}}$\Omega$\\ {[}l{]}\end{tabular}} & \textbf{\begin{tabular}[c]{@{}c@{}}$t_{rest}$\\ {[}s{]}\end{tabular}} & \textbf{\begin{tabular}[c]{@{}c@{}}$t_{min}$\\ {[}s{]}\end{tabular}} \\ \midrule
\begin{tabular}[c]{@{}l@{}} Unconfined \\ compression\end{tabular} & yes & $\frac{\rho gH}{\sqrt{3}}$ & - & $140$ & $0.54$ & $10-100$ & $1-10$ \\
Slug test & yes & $\sqrt{\frac{\rho g \mu_p V}{1.074}}$ & - & $\frac{\tau{D}\sqrt{3}}{\rho g}$ & $\frac{S\tau{D}\sqrt{3}}{\rho g}$ & 0 & $\frac{\Omega}{Q}$ \\
Abrams cone & yes & - & $\frac{\rho gH}{\sqrt{3}}$ & $304$ & $5.5$ & \textcolor{red}{60} & $1-10$ \\
Mini cone & yes & - & $\frac{\rho gH}{\sqrt{3}}$ & $150$ & $0.67$ & \textcolor{red}{60}  & $1-10$ \\
penetrometry & no & - & - & - & $\simeq 1$ & $10-100$ & - \\ \bottomrule
\end{tabular}
\caption{Popular tests for yield stress evaluation and their characteristics. Dashes indicate machine-dependent parameters.}
\label{tab:test map}
\end{table}

The following of this section discusses some aspects to consider when handling materials with a fast structuration.

\subsection{Serial versus continuous measurement} 

\begin{figure}[htbp] 
    \centering 
    \includegraphics[width=0.7\columnwidth]{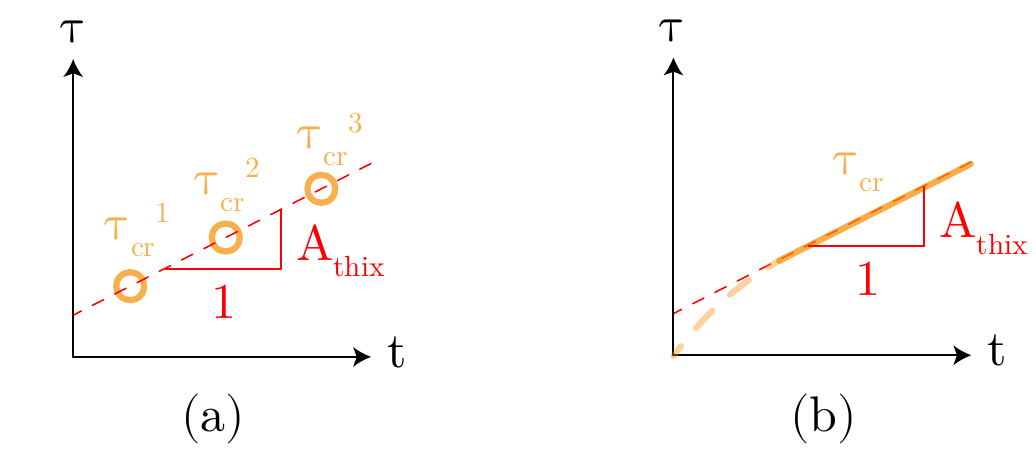} 
    \caption{Diagram of principles of serial (a) and continuous (b) measurement of the structuration rate, assumed linear at an early age and described by a single parameter $A_{thix}$.} 
    \label{fig:athix_discrete_vs_continuous} 
\end{figure}

The structuration rate is measured through two distinct approaches. The first one consists in carrying out a serial measurement, i.e. a series of measurements of the yield stress $\left(\tau_{cr}^n\right)$ at different ages $t_n$ and on distinct samples (Figure \ref{fig:athix_discrete_vs_continuous}a). In the case of cement mortars and yield stress fluids in general, these measurements can be carried out without a rheometer, thanks to practical techniques whose principle is to trigger the material flow or rupture by destructive mechanical stress. Many so-called workability tests, which aim to qualify the flow properties of concrete empirically, are based on this principle \cite{bartos_workability_2002} but only some allow to measure yield stress using appropriate physical modeling.  
The Abrams cone test uses a model proposed in \cite{roussel_fifty-cent_nodate} and there are also techniques from soil mechanics (shear box \cite{assaad_measurement_2014}, rapid indentation) or from industry (squeeze test \cite{toutou_squeezing_2005} also called \textit{squeezing test}) which provide satisfactory measurements of mortars and cementitious pastes yield stress. Vane shear test can also be used for the evaluation of yield stress of 3d printing mortars \cite{RAHUL201913}. \\
The second approach, called continuous, consists in continuously characterizing $\tau_{cr}$ by a single measurement extended over time (Figure \ref{fig:athix_discrete_vs_continuous}b) and carried out on a single sample. Slow penetrometry (review in \cite{reiter_role_2018}) corresponds to this typology. 
The two approaches have their merits: the serial measurements only provide snapshots of the material properties, but they may be easier to replicate on several samples and provide thus statistical information such as confidence intervals or might give access to mix homogeneity. Continuous measurements provide more granularity, but usually rely on more expensive equipment and are done once, thus preventing statistical analysis.

\subsection{Benchtop versus inline measurement}

The characterization of the build-up rate can be carried out on the bench or inline, i.e. either beforehand by producing the material and carrying tests with generic laboratory equipment, or directly during printing.\\ 
The benchtop measurements have the advantage of being reproducible with simple tools (table mixer, balance...) but do not guarantee that the studied material is exactly the one printed, as shown in this paper.
The inline test can be realized more easily on the very new printed mortar and can be realized many times at the same state of structuration because the accelerated material is produced continuously. Moreover, it has the great advantage to characterize the material that has gone through the same history of solicitation as the actual printed material. 
From the previous review, we can conclude that in situ easy-to-perform tests and corresponding methodology needs to be developed to measure efficiently the yield stress at different time frame using apparatus that can cover many decades to measure the build-up rate of the material and be able to set properly the printing parameter during a printing session. The next subsections precise some last definitions.

 \subsection{Measurement window}
 Material tests usually have a specific operating domain and can measure values of yield stresses $\tau_{cr}$ within a window $[\tau_{min},\tau_{max}]$. When the build-up rate is high, the time interval $[t_{max}-t_{min}]$ during which the series of tests can be carried out, later called \textit{measurement window}, is generally small. 

\label{subsection:measurement_window_cone} 
\begin{figure}[htbp] 
    \centering 
    \includegraphics[width=0.9\columnwidth]{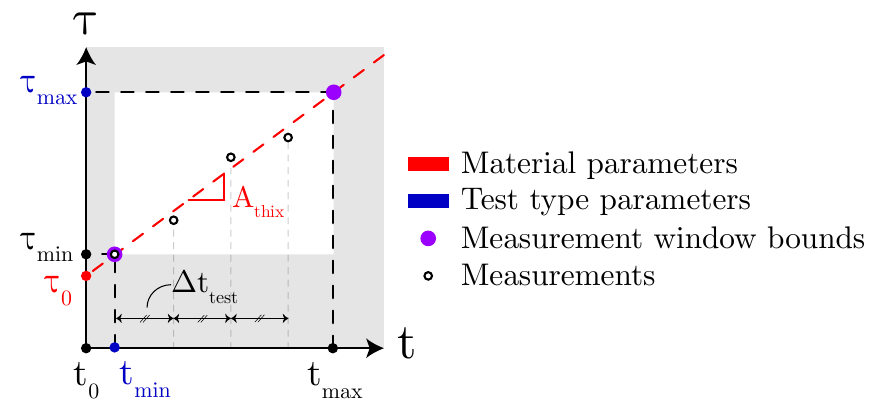} 
    \caption{Graph illustrating measurement window bounded by the points $(t_{min},\tau_{min})$ and $(t_{max},\tau_{max})$.}
    \label{fig:measurement_window} 
\end{figure} 

Figure \ref{fig:measurement_window} illustrates a typical measurement window and highlights that some parameters depend on the material, $A_{thix}$ and $\tau_{0}$, and others, $t_{min}$, $\tau_{max}$ depend on the test type. For a two-component 3D printing, large $A_{thix}$ reduces drastically $t_{max}$ and thus the size of the measurement window. \\

\subsection{Number of possible tests inside the measurement window} 
\label{subsection:test_delay_paillasse} 
Another important point is to anticipate how many tests are possible inside the measurement windows. Considering, that measuring $A_{thix}$ by a linear regression requires $n$ trials, at least $n=3$, and $\Delta t_{test}$ the delay between 2 successive measurements, the number $n$ of possible tests verifies  the following inequality (Eq. \ref{eq:tmintmax_vs_nmax}):

\begin{equation} 
t_{max} - t_{min} \geq n \cdot \Delta t_{test} 
\label{eq:tmintmax_vs_nmax}
\end{equation} 
 
The spacing delay between trials $\Delta t_{test}$ will be at least equal to $t_ {min}$, which corresponds to the duration of the test. However, depending on the precision $\delta_\tau$ of the type of test, it may be necessary to increase this delay because $\Delta t_{test}$ must be large enough to be able to obtain sufficiently contrasted measurements. From Eq.\ref{eq:athix}, the uncertainty/structuration rate ratio $\frac{\delta_{\tau}}{A_{thix}}$ gives a minimum value for $\Delta t_{test}$: 

\begin{equation} 
\Delta t_ {test} \geq \frac{\delta_{\tau}}{A_{thix}} 
\label{eq:delta_t_test} 
\end{equation}

\subsection{Homogeneity}
\label{homogeneity}
The homogeneity of the material across the tested sample is another significant aspect to keep in mind designing a test. When the material has a rapid structuration, the yield stress evolves during the preparation of the sample. From Eq.\ref{eq:athix}, we write $\Delta \tau_{cr}$ the variation of yield stress across the sample during $t_{sample}$.
\begin{equation}
    \Delta \tau_{cr} \sim A_{thix}\cdot t_{sample}=\frac{A_{thix}\Omega}{Q}
    \label{eq:delta_tau}
\end{equation}

The material is tested after a given time $t^*$.
A necessary condition for material homogeneity is that the variation of yield stress across the sample remains small compared to the measured yield stress:
\[
    \frac{\Delta \tau_{cr}}{\tau_{cr}\left(t^*\right)}=\frac{A_{thix}\cdot t_{sample}}{A_{thix}\cdot t^*+ \tau_0}=\frac{t_{sample}}{t^*+ \frac{\tau_0}{A_{thix}}} \ll 1
\]
Introducing $t_{thix}=\tau_0/A_{thix}$ which is a characteristic time of the material, corresponding (see Eq.\ref{eq:athix}) to the time need for the material to double its initial yield stress, the
condition for a good homogeneity is \eqref{eq:homogeneity}: 
\begin{equation}
    t^* \gg t_{sample}-t_{thix}
    \label{eq:homogeneity}
\end{equation}
Having in mind these different parameters and constraints regarding window measurement size, number of possible tests, and homogeneity conditions, we propose in the following, original method using a so-called pocket hand vane test. The device, the methodology, and the first results are detailed.

\section{Pocket hand vane presentation and mechanical analysis} 

\label{sec:scissometer_insitu}
\begin{figure}[h!] 
    \centering 
    \includegraphics[width=0.4\columnwidth]{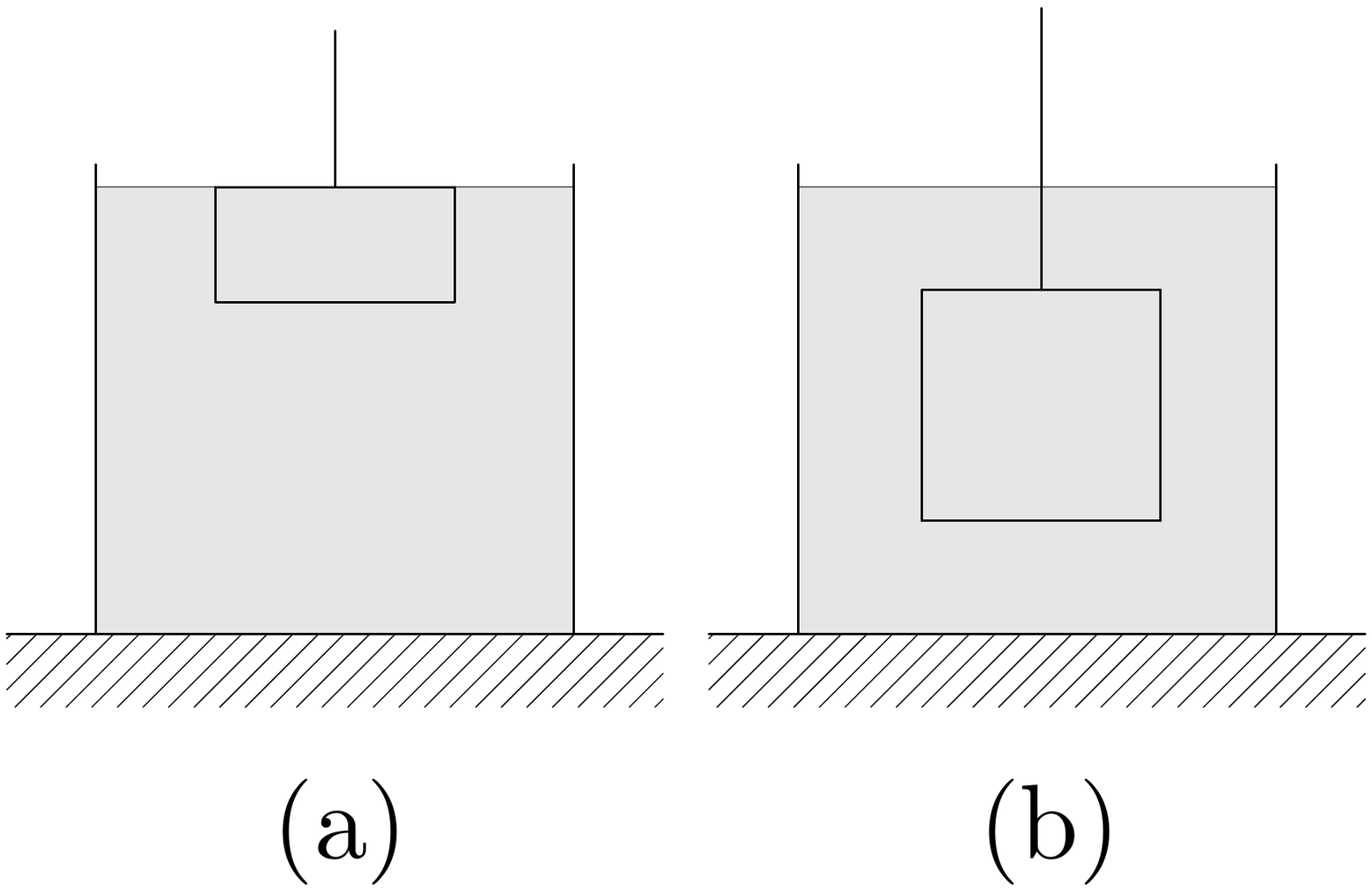} 
    \caption{Diagram of the measurement area with a pocket vane (a) and the classic vane (b).} 
    \label{fig:torvane_vs_vane} 
\end{figure}

The principle of the vane test (or scissometer) is to shear a sample of material by applying a torque to it while measuring the stress necessary to trigger its flow or its rupture. Just like the cone test, it prescribes a perfectly known stress field, here a pure shear. This type of measurement on a bench with an immersed vane, mobilizing fairly large quantities of mortar is classic \cite{astm_standard_2019} (Figure \ref{fig:torvane_vs_vane}b) to measure the shear yield stress\cite{austin_workability_1999} and structuration rate \cite{roussel_distinct-layer_2008} of cementitious materials dedicated to casting, and the results obtained can be accurately correlated with a rheometer \cite{omran_portable_2011}.

\subsection{The pocket vane}
The pocket vane (also called \textit{Torvane}) is a smaller version of the scissometer, except that the vane is positioned on the surface of the specimen (Figure \ref{fig:torvane_vs_vane}a). Initially developed for fine-grained soils \cite{astm_standard_2019}, it has never been used, to our knowledge, on cementitious materials. Like a hand vane, it has a knurled head that is manually turned and connected to a vane shaft. But, the loading (i.e. shear stress proportional to the torque) is localized in the first few millimeters at the surface of the specimen.  
The exerted stress value can be related to the value of the torque given by the graduations on the apparatus head, thanks to the analysis made in the next section. 

\begin{figure}[htbp] 
    \centering 
    \includegraphics[width=0.7\columnwidth]{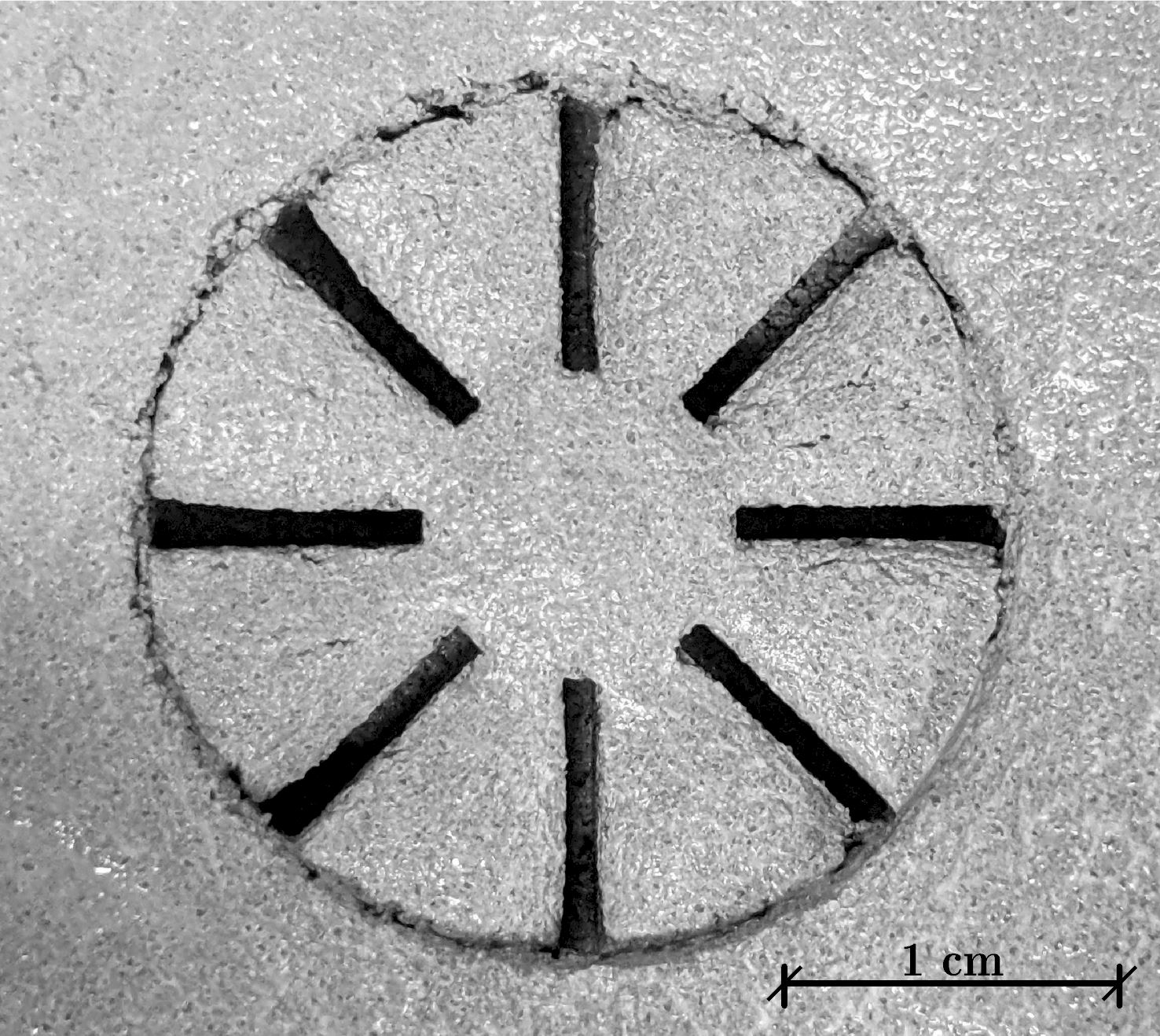}
    \caption{The test area of a printable mortar tested with a pocket vane with the Standard vane head at a very young age. The imprint of the vanes and the cylindrical rupture is clearly visible.} 
    \label{fig:zone_cisaillement_scisso} 
\end{figure} 

\subsection{Mechanical analysis}
The mechanical analysis was initially proposed for the determination of undrained cohesion of clays studied in geotechnics \cite{cadling_vane_1950} and considers that the flow stress is exerted on a cylinder inscribed around the blades. Indeed it has been shown that at the moment of flow, the cylinder of material embedded in the blades rotates like a rigid body. The flow is located uniformly along a thin cylindrical layer near the tips of the blades \cite{keentok_shearing_1985}, clearly visible in Figure \ref{fig:zone_cisaillement_scisso} .\\
For a classic scissometer, the measured torque $T_{cr}$  is due to two components, one resulting from the shear on the lateral curved edge $T_{s}$ and the other $2 T_{e}$ from the shear on the two upper and lower horizontal edges of the inscribed cylinder  \cite{dzuy_direct_1985} \cite{boger_rheology_2000}: 

\begin{equation} 
T_{cr} = T_{s} + 2 T_{e} 
\label{eq:shear_equilibrium_scissometer} 
\end{equation} 

The case of the pocket shear vane is different. The blades are inserted at the surface so there is only one sheared horizontal edge: 

\begin{equation} 
T_{cr} = T_{s} + T_{e} 
\label{eq:shfine-grainedium_pocketscissometer } 
\end{equation} 

 \begin{figure}[h!]
    \centering 
    \includegraphics[width=0.7\columnwidth]{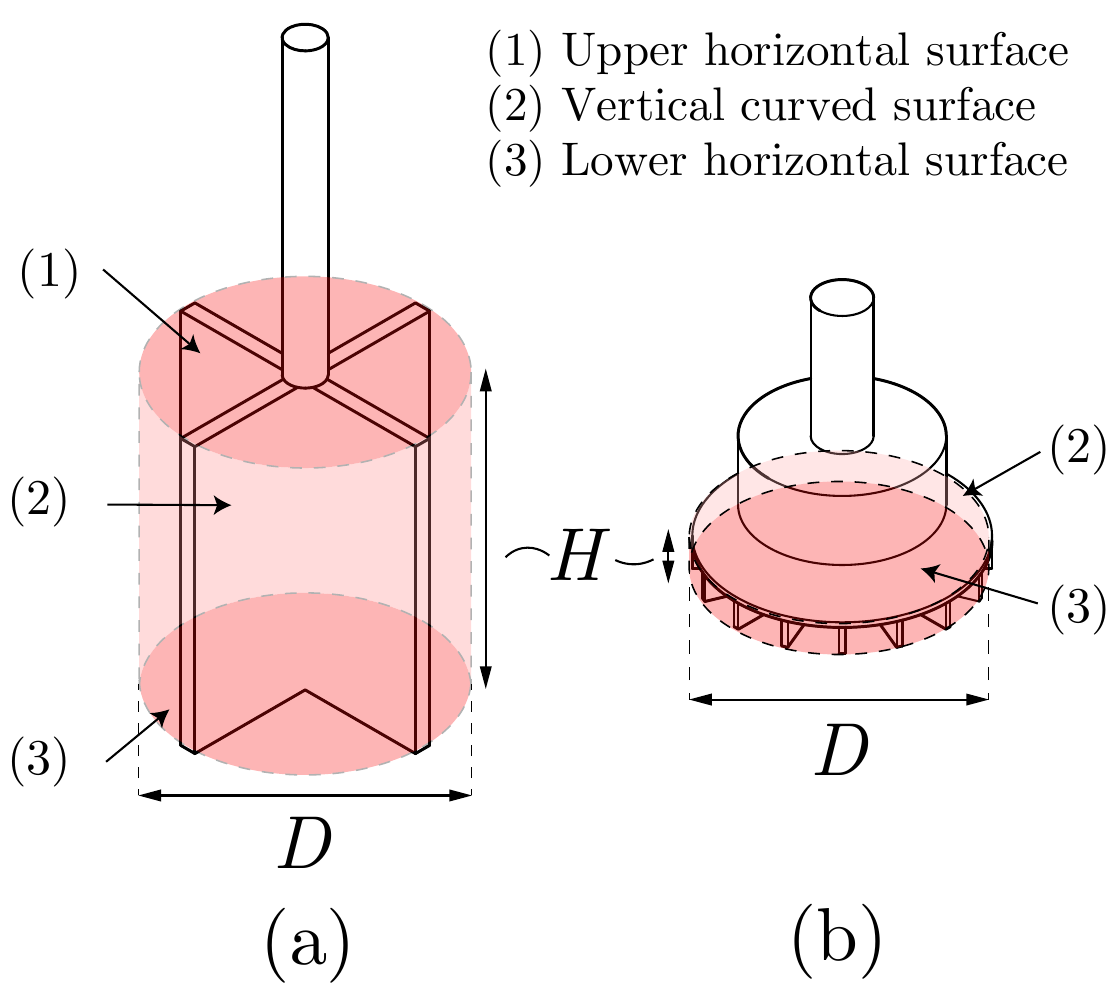}
    \caption{Diagram of sheared surfaces with a classic vane (a) and a pocket vane (b).} 
    \label{fig:scisso_zones} 
\end{figure} 

These differences are illustrated in Figure \ref{fig:scisso_zones}. 
$T_{s}$ is given by :  

\begin{equation} 
T_{s} = \left(\pi DH \right) \frac{D}{2} \tau_{s} 
\end{equation}
where $\pi DH$ represents the curved surface of the cylinder, with $D$ the diameter of the cylinder, $H$ its height, and $\frac{D}{2}$ the lever arm. 

$\tau_{e}(r)$ is not known a priori, probably linear in $r$, and the balance of the torque exerted during the flow \ref{eq:shear_equilibrium_pocketscissometer} is: 

\begin{equation} 
T_{cr} = (\pi DH) \frac{D}{2} \tau_{s} + 2 \pi \int_0^{D/2} \tau_{e} (r) r^2 dr 
\label{eq:shear_equilibrium_pocketscissometer} 
\end{equation} 

At the time of the flow, it is assumed that the constraints $\tau_{s}$ and $\tau_{e}$ are uniform and equal to $\tau_{cr}$. In this case, the Equation \ref{eq:shear_equilibrium_scissometer} is reduced to a simpler form where we can directly express $T_{cr}$ according to the dimensions of the blades and $\tau_{cr}$:

\begin{equation} 
T_{cr} = \frac{\pi D^3 \tau_{cr}}{12}\left(1 + \frac{6H }{D}\right)
\end{equation} 

It is noteworthy that the test prescribes a pure shear stress state, with a negligible elongational or compressive component, unlike other types of tests (penetrometry, Abrams cone test).

\section{Material and protocols}
We define now the used material and the protocols for the different experimentations that we carry out, benchtop and inline experimentations.
\subsection{Material and preparation}
In this study, the mortar is formulated using the 3DPG dry mix provided by Lafarge. The maximum particle size of this dry mix is $\phi_{max}=0.8\mathrm{mm}$. The water (including the superplasticizer water) to powder mass ratio is equal to $\frac{E}{P}=0.1$. The phosphonates base superplasticizer is adjusted to reach an initial yield stress comprised between 100 and 200 Pa. The resulting superplasticizer quantity is usually comprised between 0.5\% and 0.6\% of the dry mix mass.
Then, an aluminum sulfate-based accelerator Floquat ASL is added varying the dosage between 1.5g/kg and 17g/kg (1 and 17 mL/kg) to obtain different build-up rates.

The mixing process is either made using a benchtop mixing unit, the Hobart HSM10 mixer, or the mixing unit of the XtreeE printing cell.

For both production the mixing protocol is the following: 
\begin{enumerate}[label=\alph*)]
\item the superplasticizer and the water are weighted and mixed in one container
\item The dry mix is weighted and introduced in the mixing unit
\item The mixing unit is started (at low speed for the benchtop unit) and the liquids are slowly poured for approximately 1 min. 
\item the mixing is continued for 5 more minutes (at high speed when using the benchtop unit)
\end{enumerate}

\begin{figure} [h] 
    \centering 
    \includegraphics[width=1\columnwidth]{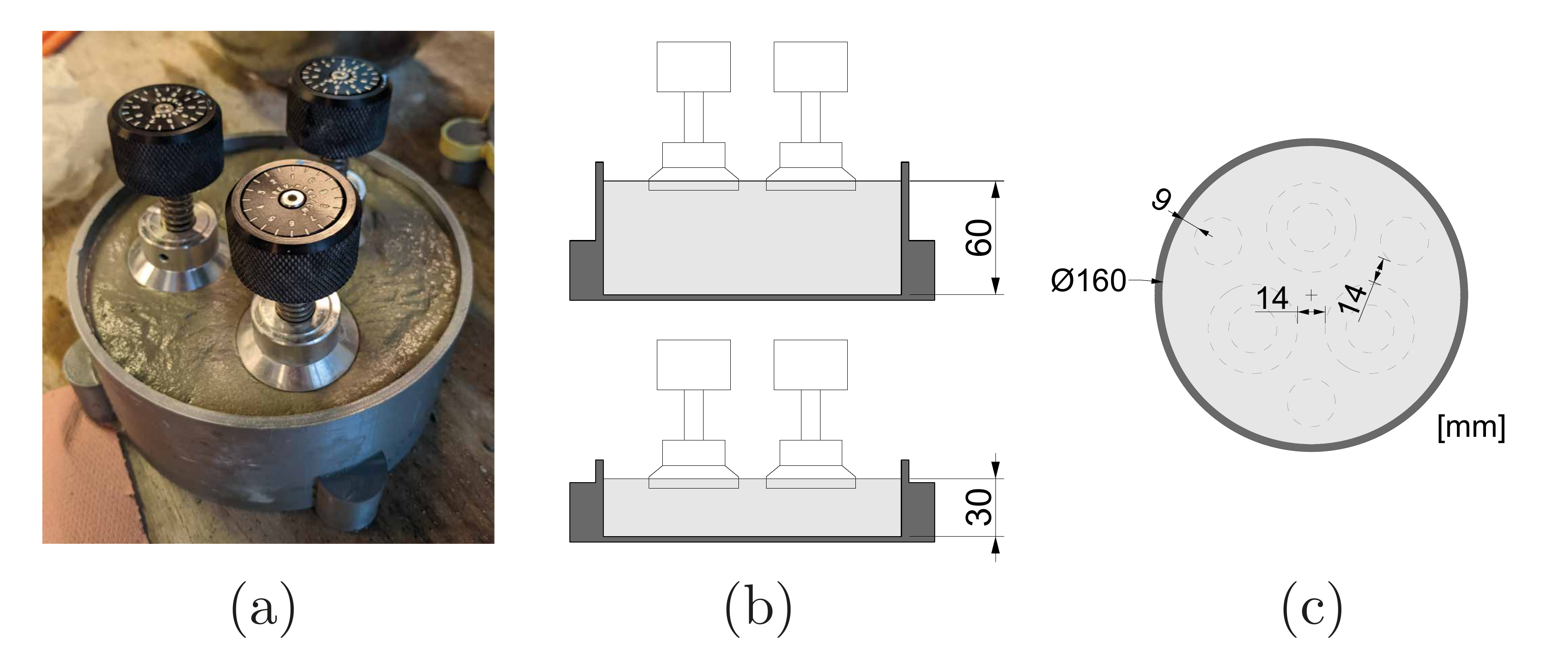}
    \caption{The ex-situ measuring device with three pocket scissometers, (b) two container heights are tested, (c) top view of the container and position of the 2 sets of 3 tests in dotted line.} 
    \label{fig:scissometers_resume} 
\end{figure}

Concerning the accelerator, for the benchtop it is added and dispersed following this protocol:
\begin{enumerate}[label=\alph*)]
\item the accelerator solution is weighted in a small recipient
\item the mixing unit is started at a large speed
\item the accelerator is poured quickly and the chronometer is started (t=0)
\item The mixing is continued for 30 more seconds
\end{enumerate}
For the inline test, the accelerator is added through the printing head via a dosing micro-pump and following the XtreeE process.\\
After the end of the acceleration protocol, the accelerated mortar is for both tests (benchtop and inline), poured into containers with different sizes (see Figure \ref{fig:scissometers_resume}).
In the following, \textit{ex-situ} tests refer to these tests realized in containers. Note that the free surface of the container needs to be as smooth as possible by tapping by hand or using a spatula. This facilitates the insertion of the scissometers into the material. \\
For the inline situation, dedicated printing laces are also carried out for direct \textit{in-situ} tests. \\The build-up rate measurements use the devices and protocols described in the next section. 

\subsection{Tests protocols}

The pocket vane used in this paper is the Humboldt H-4212MH that refers to the ASTM D8121/D8121M standard \cite{astm_standard_2019}. 

 \begin{figure}[h!]
    \centering
    \includegraphics[width=0.3\columnwidth]{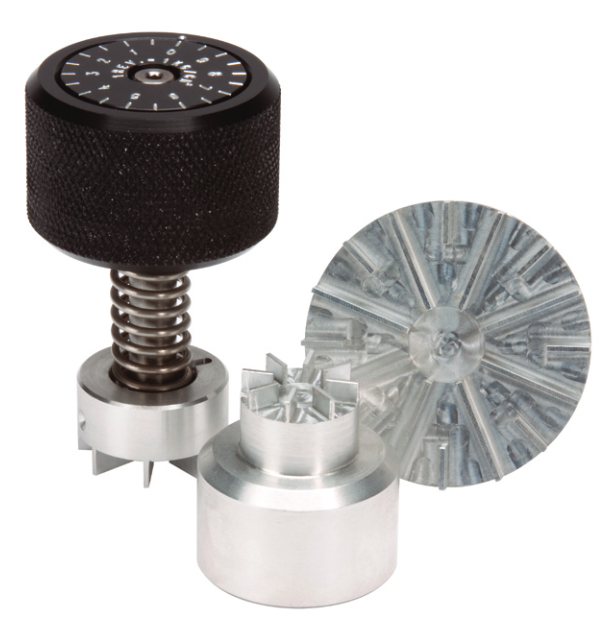}
    \includegraphics[width=0.4\columnwidth]{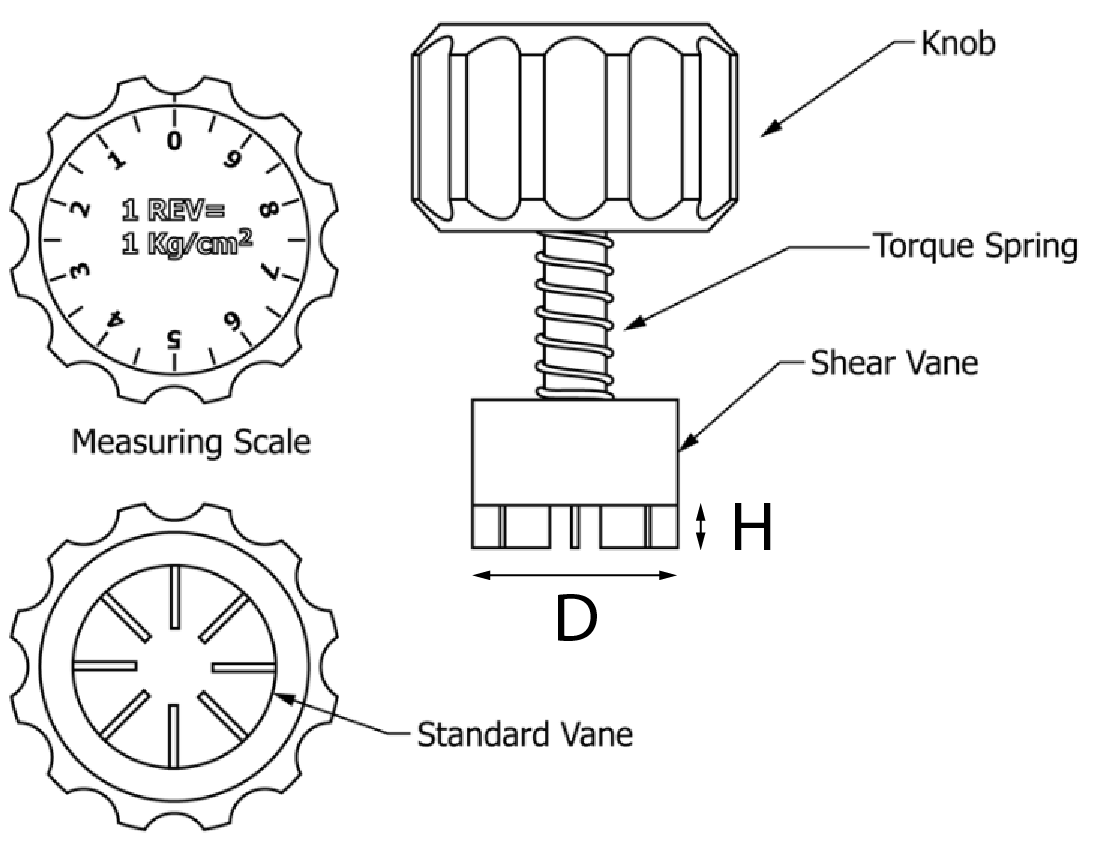}
    \caption{Left : a pocket shear vane \textit{Humboldt H-4212MH}, right : schematic from ASTM D8121/D8121M standard \cite{astm_standard_2019}.}
    \label{fig:illustration_torvane}
\end{figure}

\begin{table}[]
\centering
\begin{tabular}{|l|l|l|l|}
\hline
Vane head & $D$ (mm) & $[\tau_{min}, \tau_{max}]$ (kPa) & $\delta_{\tau}$ (kPa) \\ \hline
Sensitive        & 47.5 & 1 - 20  & 1  \\ \hline
Standard & 25.5 & 5 - 100 & 5 \\ \hline 
Large Capacity & 19 & 12.5 - 250 & 12.5 \\ \hline 
\end{tabular} 
    \caption{Data relating to the different vane heads of the pocket shear vane. $D$: Diameter of the sheared surface, $[\tau_{min}, \tau_{max}]$: Yield stress measurement domain (kPa), $\delta_{\tau}$: Accuracy of the graduation. } 
    \label{table:calibre_scisso} 
\end{table}

Three interchangeable vane heads are available, which are chosen according to the strength range of the studied material and give access to different decades of yield stress values. The data are presented in Table \ref{table:calibre_scisso}. The measuring range increases as the diameter of the sheared surface $D$ decreases. The blade height $H$ is constant, 5.2 mm. The maximum torque applicable using the instrument corresponds to a complete rotation of the head (10 graduations).
The change in diameter of the blades has a limited influence on the measurement of fine-grained soils \cite{dzuy_direct_1985}, this will be discussed later in this article.

The yield stress evolution is then measured using this apparatus and the appropriate vane geometry. The vanes are inserted in the samples and tests are carried out by turning each vane at a constant low speed (approximately 3 rpm) until peak stress is reached.

\subsubsection{Ex-situ benchtop and  ex-situ inline tests }
\begin{figure}[htbp] 
    \centering 
    \includegraphics[width=0.9\columnwidth]{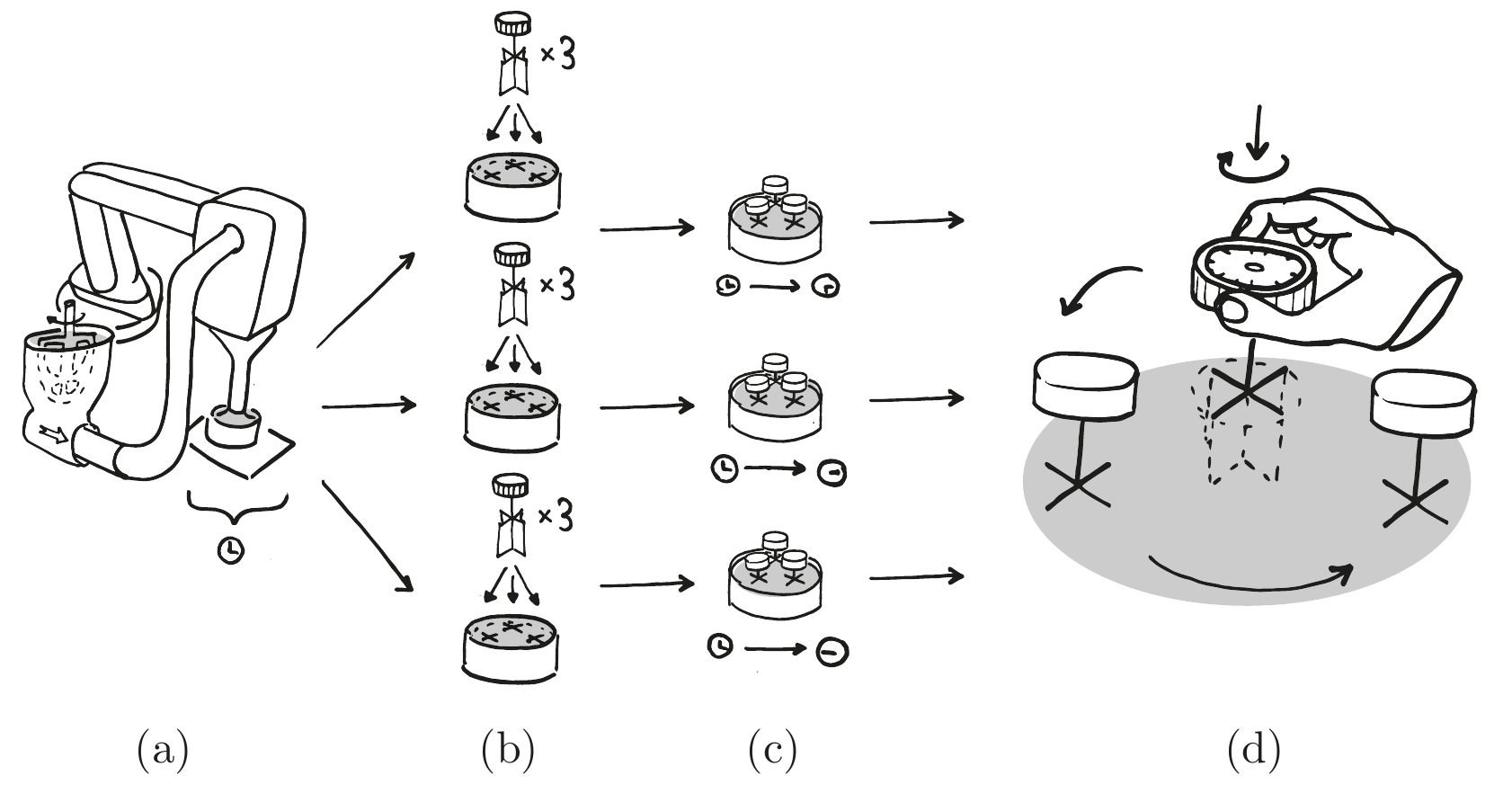}
    \caption{Illustration of the steps of an ex-situ inline vane test: (a) Preparation of the sample: filling a container with the printed material (b) Inserting the vanes or footprints (c) Waiting for a given amount of time  (d) Performing sets of three tests for each sample. } 
    \label{fig:insitu_test_storyboard} 
\end{figure}

For the ex-situ tests the scissometers or footprints are placed following the arrangement shown in Figure \ref{fig:scissometers_resume} c.
The footprint refers to a plastic model of the different vane geometry made by FDM 3D printing process. Inserted just after the container is filled, when the mortar is still fresh, they may help to introduce the vane head for testing without damaging it. These footprints are replaced with the real vane apparatus just before testing.
Sets of 3 tests are carried out to obtain an average value and dispersion estimation at each time and each corresponding yield stress value. By rotating $\pi/3$, it is thus possible to successively carry out 2 sets of 3 tests per container: the first set with the sensitive vane head for a very young material, and the second one with the standard vane head for a more structured material (see Figure \ref{fig:insitu_test_storyboard}).\\ 

\subsubsection{In-situ inline tests}
\begin{figure}[h] 
    \centering 
    \includegraphics[width=0.4\columnwidth]{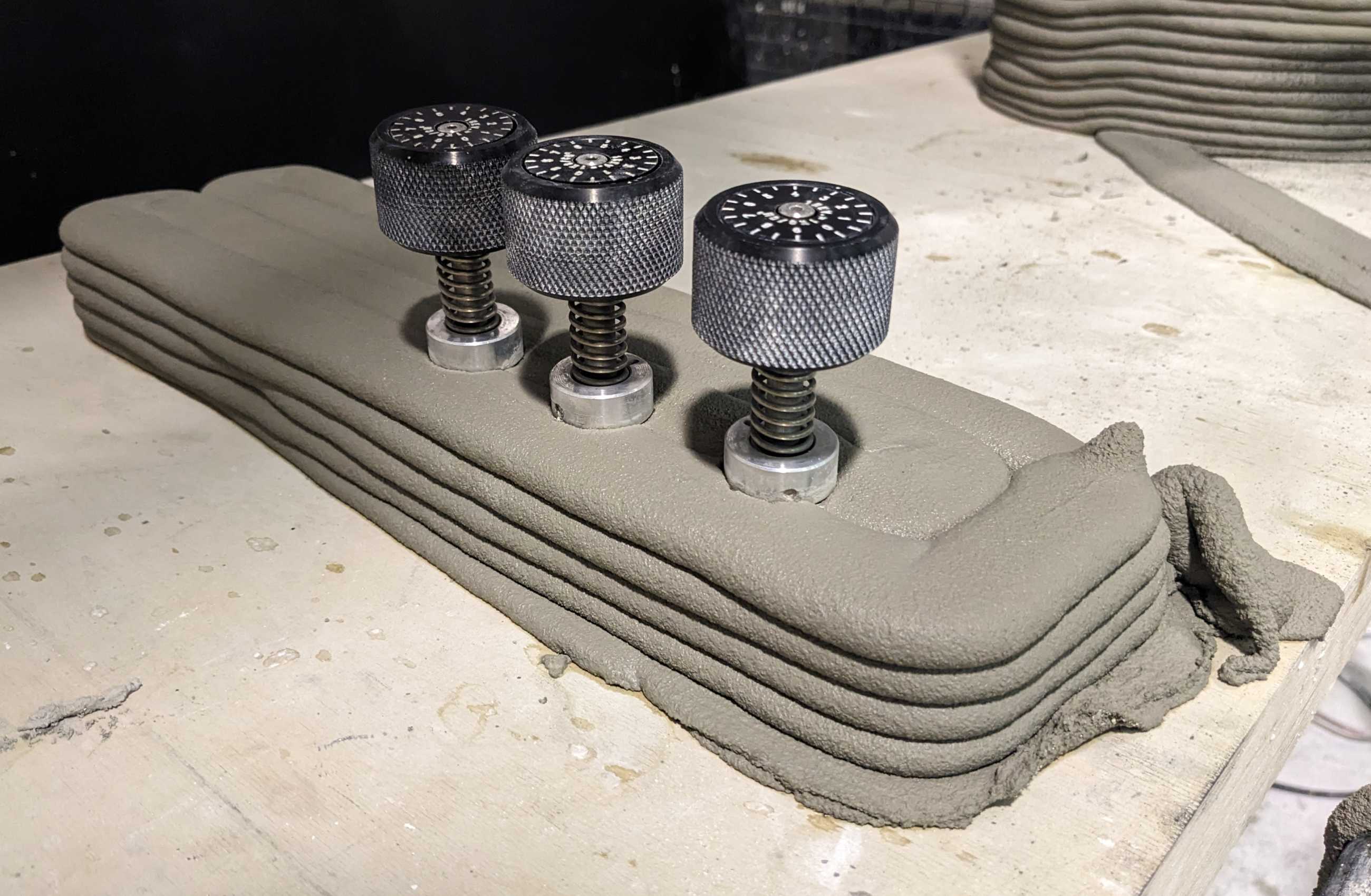} 
    \includegraphics[width=0.4\columnwidth]{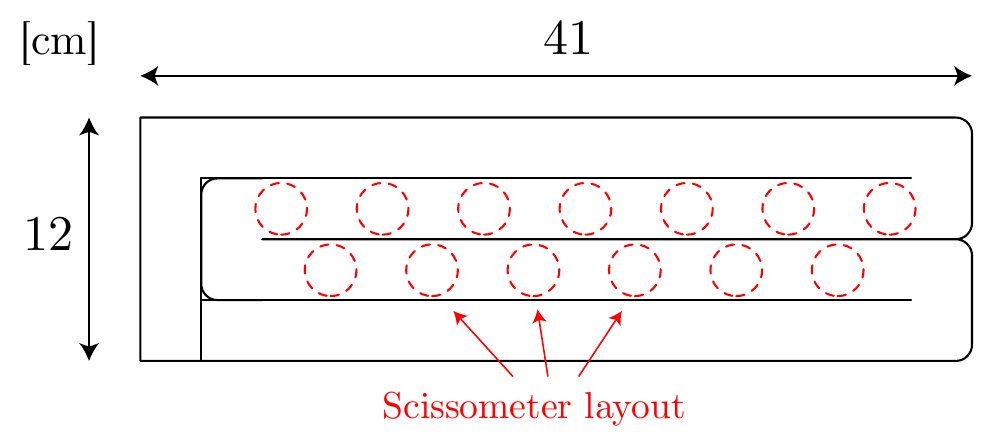} 
    \caption{Illustration of an in-situ inline vane test : Left: a printed sample dedicated to the lace measurement. Right: Layout of vanes on the printed sample.} 
    \label{fig:scisso_boudins} 
\end{figure}

For in-situ inline tests, meaning direct tests on printed laces, the protocol is almost the same, but directly along the top of the printed wall (fig.\ref{fig:scisso_boudins}).

The Standard vane is chosen (diameter 25.5 mm) and the lace width is set to be around 30 mm, a common value with our system. 
To avoid any edge effect, the vanes will be arranged according to the diagram shown in Figure \ref{fig:scisso_boudins}.
The yield stress is also measured at different resting times for similar accelerator content as for the ex-situ inline test. 

\section{Preliminary tests}
In this section, preliminary tests are investigated to judge the relevance of the proposal.
\subsection{Comparison of classical and pocket scissometer benchtop measurements} 

 \begin{figure}[h!] 
    \centering 
    \includegraphics[width=0.4\columnwidth]{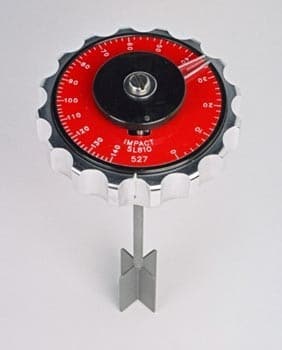} 
    \caption{Photograph of a classic scissometer (Source: mtlabs.co.nz)} 
    \label{fig:img_classic_scisso} 
\end{figure} 
To validate the tool and the operating mode, first benchtop experiments are carried out. A comparison between the yield stress measurements made using a standard hand vane (Figure \ref{fig:img_classic_scisso}), usually used by the company \textit{XtreeE} and already qualified by the community for yield stress measurement, and the pocket hand vane, are made. In this case, the accelerator dosage was chosen equal to 7.5g/kg of dry mix. 

\begin{figure}[h!]\includegraphics[scale=0.4, angle=0]{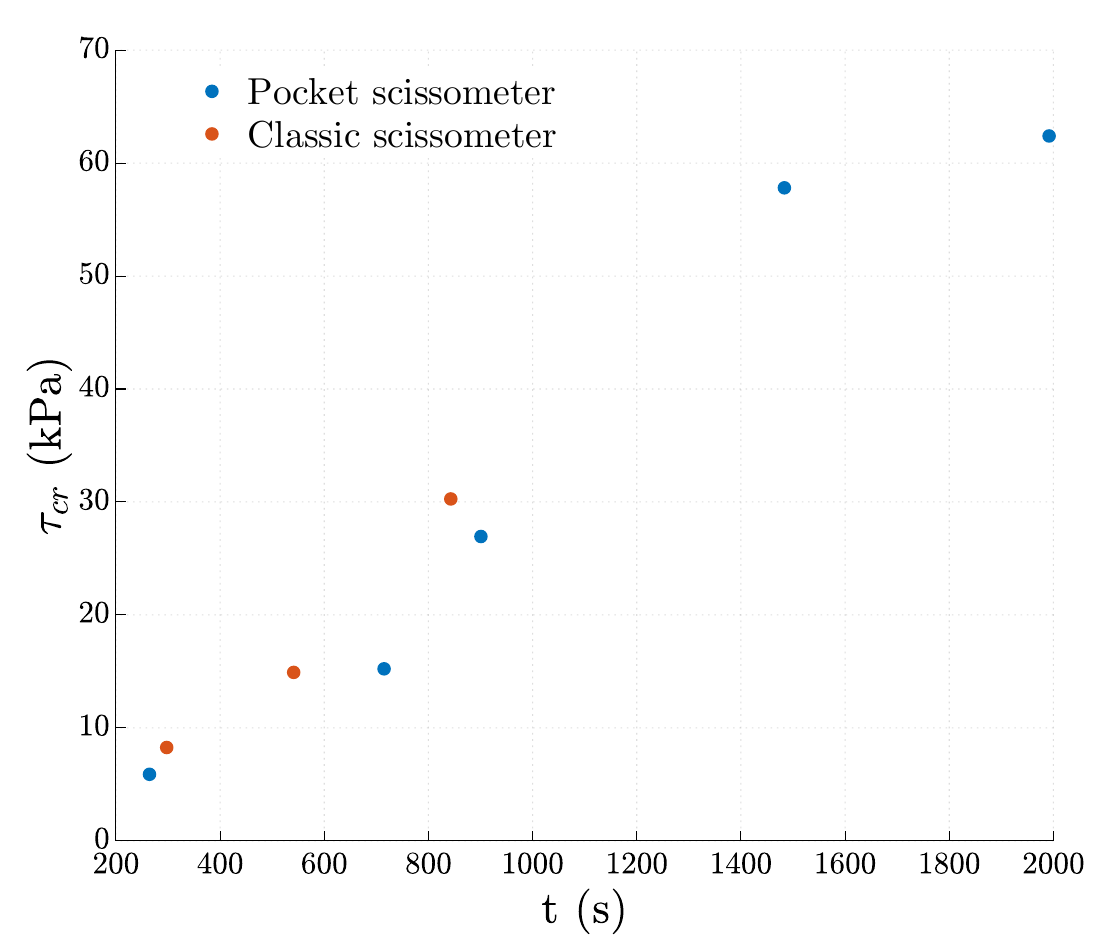} 
    \centering 
    \caption[]{Benchtop measurements of shear yield stress $\tau_{cr}$ at different resting time $t$ of the same mortar, with pocket and classic vane.} 
    \label{fig:vane_vs_torvane_results} 
\end{figure}

The vane measurements are made on numerous prismatic containers filled after the dispersion of the accelerator in the benchtop mixing unit. After given resting times, the hand vanes are inserted inside the tested specimen and slowly rotated at a  similar speed ($\approx 3$ rpm). The results are reported in Figure \ref{fig:vane_vs_torvane_results} and are comparable, which assesses the correct precision of the absolute yield stress measured value. Moreover, the evolution measured by both apparatus is also really similar which demonstrates, first the assumed linear evolution (Eq.\ref{eq:athix}), and the potential of the pocket vane geometry to measure the build-up rate of fine-grained printable materials.

\subsection{First campaign of benchtop measurements using a pocket scissometer} 

A series of benchtop experiments are carried out to estimate better the accuracy that can be expected from the pocket vane and verify the hypothesis taken in the theoretical analysis. 
A total of 10 series were carried out for increasing dosages of accelerator $D_{acc}$ (9, 13, 17 g/kg). 
To assess possible wall effects, two container depths (30 and 60mm) were tested (Figure \ref{fig:scissometers_resume}b), yielding comparable results.

As mentioned before sets of 3 tests are carried out to obtain an average value at each time. 
The results are presented in Table \ref{table:results_scisso_lab}. A total of 10 series were carried out for increasing dosages of accelerator $D_{acc}$ (6, 9, 12 mL/kg). $A_{thix}$ is deduced from a linear regression from Eq. \ref{eq:athix} and the average residual $\bar{R}$ of the measurements compared to the model is given.\\ 
The calculated $A_{thix}$ are well repeatable and in the expected order of magnitude with an experimental uncertainty $\bar{R}$ similar to the uncertainty $\delta{\tau}$ of the instrument for each vane head indicated in the Table \ref{table:calibre_scisso}.
At 12mL/kg, we reach the measurement limits of the instrument, the material being too structured. Note also the importance of the proposed prior footprint: for tests $n^{o}$ 6 and 9, underlined in Table \ref{table:results_scisso_lab},  damages due to blades introduction, artificially decrease the yield stress values.\\ 
The relevance of the tool being established, we propose here its use inline.

\begin{table}[] 
\begin{tabular}{|r|l|l|r|r|r|} 
\hline 
\multicolumn{1}{|l|}{$i_{exp}$} & 
  Footprints & 
 Vane head(s) & 
  \multicolumn{1}{l|}{$D_{acc}$(mL/kg)} & 
  \multicolumn{1}{l|}{$A_{thix}$(Pa/s)} & 
  \multicolumn{1}{l|}{$\bar{R}$(Pa)} \\ \hline 
1 & no & Sensitive & 6 & 20.3 & 1373 \\ \hline
2  & no                                 & Sensitive            & 6  & 20.2                       & 1029                        \\ \hline
3  & no                                 & Sensitive            & 6  & 16.5                      & 1453                        \\ \hline
4  & no                                 & Sensitive + Std & 6  & 21.6                        & 1452                          \\ \hline
5  & no                                 & Sensitive + Std & 6  & 20.8                       & 3348                          \\ \hline
\underline{6}  & {\color[HTML]{FF0000} \underline{\textbf{no}}} & \underline{Standard} (Std)            & \underline{9}  & {\color[HTML]{FF0000}  \underline{54.2}} & {\color[HTML]{FF0000} \underline{6533}} \\ \hline
7&yes&Standard&9&69.9&2989\\\hline 
8&yes&Standard&9&89.6&3003\\\hline 
\underline{9}&{\color[HTML]{FF0000}\underline{\textbf{no}}}&\underline{Standard}& \underline{12} & {\color[HTML]{FF0000} \underline{44.1}} & {\color[HTML]{FF0000}\underline{30000}} \\ \hline 
10 & yes & Standard + HC & 12 & 67.3 & 5393 \\ \hline 
\end{tabular} 
\caption{Experimental benchtop results with the pocket vane. Notation: $i_{exp}$: experiment number, $D_{acc}$ dosage of the accelerating additive, $\bar{R}$ mean residual of the $A_{Thix}$ estimate.  \underline{Underlined} results for tests without prior footprints.} 
\label{table:results_scisso_lab} 
\end{table}

\section{Inline measurement on the printed material}

This section addresses the final objective of the proposal, which is to demonstrate the opportunities of the inline method. Here the material is not mixed once and for all, but accelerated at the level of the extrusion head and therefore produced identically and continuously (unless the accelerator is not well dispersed which would be clearly visible because the quality of the print would be strongly
degraded). If the initial non-accelerated batch (40 liters) remains within its open time (approximately 45 min), each sample is produced and tested at a given age independently of the others, allowing for better precision of each yield stress assessment by increasing the number of tests made at each given resting time $t$. The spacing between trials $\Delta t_{test}$ is no longer constrained by the duration of the unit trial $t_{min}$ nor by the value of the ratio $\frac{\delta_{\tau}}{A_{thix}}$ (see Eq.\ref{eq:delta_t_test} ). 

Before the test, the initial threshold $\tau_0$ is estimated using the so-called slugs-test which consists to deduce the yield stress analytically from the weight of the mortar drops falling from the head nozzle (details in \cite{ducoulombier_slugs-test_2021}). The steps of the protocol are previously illustrated in Figure \ref{fig:insitu_test_storyboard} for the ex-situ condition (in a recipient), and in  Figure \ref{fig:scisso_boudins} for the in-situ condition, directly on a printed wall.

\subsection{Comparison between in-situ inline and 
 ex-situ inline measurements}
Let us now compare the inline results,  for tests carried out on containers filled with the robot (ex-situ, Figure \ref{fig:scissometers_resume}, or for printed samples (in-situ Figure \ref{fig:scisso_boudins}).

\begin{figure}[h] 
    \centering
    \includegraphics[width=0.6\columnwidth]{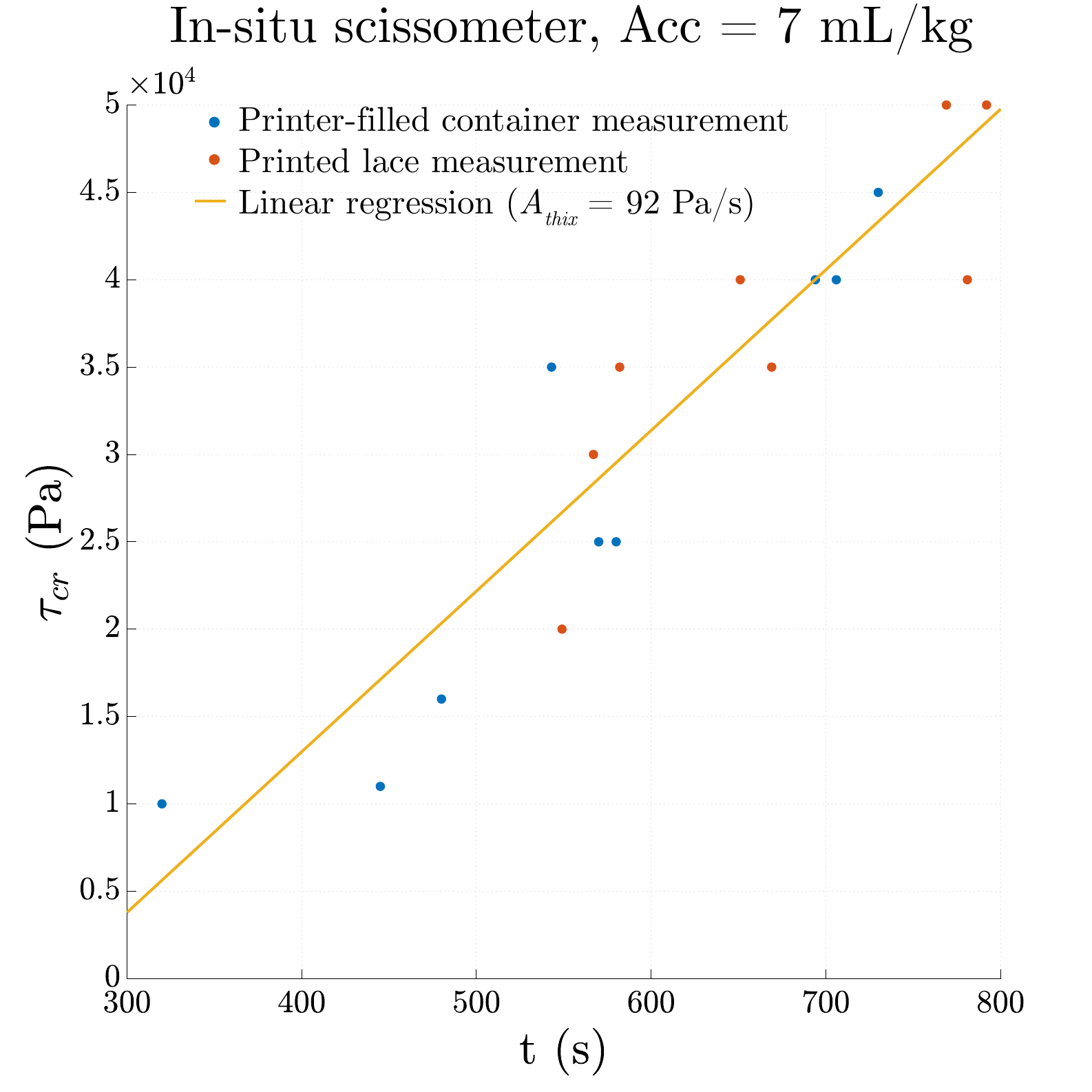} 
    \caption{inline measurements of the shear stress $\tau_{cr}$ at different ages, by measuring material from a container filled (ex-situ) with the robot or directly on printed laces (in-situ).} 
    \label{fig:scisso_boudin_vs_recipient} 
\end{figure} 

The results (Figure \ref{fig:scisso_boudin_vs_recipient}) obtained using the two methods are comparable. Indeed, all the data points (ex-situ and in-situ measurements) were correlated together and the linear regression yields a coefficient of determination $R^2$ of 0.84 which is reasonable. The correlation results of printer-filled and printed lace measurements are shown in Table \ref{table:correlation_results_scisso_boudin_vs_recipient} showing little difference in the actual structuration rate ($A_{thix}$) estimation and coefficient of determination $R^2$. Moreover, once again, the assumed linear evolution is demonstrated. In the Appendix, the tables \ref{table:appendix1_printerfilled} and \ref{table:appendix2_printedlace} provide the complete set of results.\\
In the future, repeating more of these experiments will certainly provide enough data to quantify more precisely the dispersion between the two test conditions and eventual heterogeneities.
\begin{table}[]
\centering
\begin{tabular}{|l|l|l|}
\hline
\textbf{Dataset}                      & \textbf{$A_{thix}$ (Pa/s)} & \textbf{$R^2$} \\ \hline
Printer-filled container measurements & 91                         & 0.86           \\ \hline
Printed lace measurements             & 87                         & 0.70           \\ \hline
All measurements                      & 92                         & 0.84           \\ \hline
\end{tabular}
\caption{Comparison of build-up rate between ex-situ inline measurements and in-situ inline measurements.}
\label{table:correlation_results_scisso_boudin_vs_recipient} 
\end{table}

\subsection{Comparison between benchtop and in-line measurements} 
\begin{figure}[h!] 
    \centering
    \includegraphics[width=0.8\columnwidth]{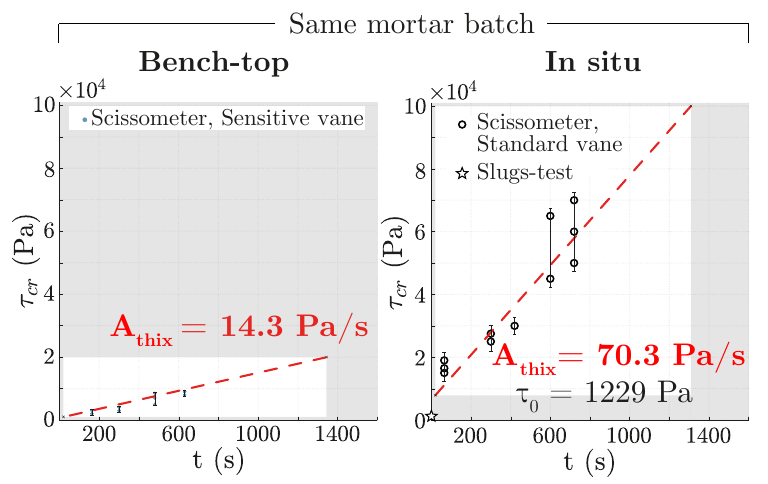}
    \caption{Time-stress graphs of inline and bench-top testing with material accelerated at a typical dosage for printing (6mL/kg). The error bars represent the interval containing the values of all measurements at a given time, plus or minus half the uncertainty $\delta_{t}$ of the vane head used.} 
    \label{fig:insitu_vs_benchtop} 
\end{figure}

This section aims to study the difference between these two approaches. The bench top test is carried out as previously described, by accelerating the mortar in a table mixer, and the fresh unaccelerated mortar is collected from the printing system so that both tests are issued from the same mortar batch.
 Figure \ref{fig:insitu_vs_benchtop} presents typical results obtained during this comparative bench-top and inline test campaigns.
 
 The trend is again linear for both test types. The dispersion at a given time demonstrates the advantage of repeating 3 times the measurements as indicated in the protocol. However, the structuration rate of the printed material is much higher than that of the benchtop: the structuring rate $A_{thix}$ of the printed material (online), $70.3 Pa/s$, is 5 times higher than that of the benchtop, $14.3 Pa/s$. Several hypotheses can be proposed:
\begin{itemize}
\item Firstly, there is the heating of the material in the pump and printing head, which may accelerate hydration reactions \cite{lothenbach_effect_2007}. 
\item Secondly, shearing inside the printing head may affect the mortar microstructure and help for good dispersion of the accelerator. 
\end{itemize}
Indeed, the shear rate experienced by the mortar within the screw auger equipped on the printing head is probably higher than that of the table mixer, resulting in better deflocculation of the material and an increase of the specific surface area of the cement grains available within the cement paste. This may allow better reactivity. \\

It is therefore clear that a bench-made material and an inline printed material do not have the same properties. This underlines the importance of inline measurements, which are more representative and allow process-related effects to be captured.

\section{Some other key figures of the method}

Here we give some information about the potential and the limits of the method.\\
Due to the benchtop procedure and the fact that the containers are small and quickly filled,  the homogeneity of the material across the tested sample (condition \ref{eq:homogeneity}) is easily verified. \\
We precise the bounds $t_{min}$, $\tau_{min}$, $\tau_{max}$ and $t_{max}$ of the measurement window of this inline test. $t_{min}$ is about 20 seconds. $\tau_{max}$ varies according to the chosen vane head: 20, 100, or 250 kPa. For $\tau_{min}$, the greater value between the one given by the Equation  \ref{eq:athix} for $t_{min}$, and the minimum threshold measurable by the vane head (Table \ref{table:calibre_scisso}), is chosen. $t_{max}$ depends on $A_{thix}$ and $\tau_{max}$ and is therefore also deduced from Equation \ref{eq:athix}. 
 $t_{max}$ may be limited not by the instrument but by the open time of the mortar, typically 45 minutes, and the volume of the batch (50L).
\begin{figure}[htbp]
    \centering 
    \includegraphics[width=0.6\columnwidth]{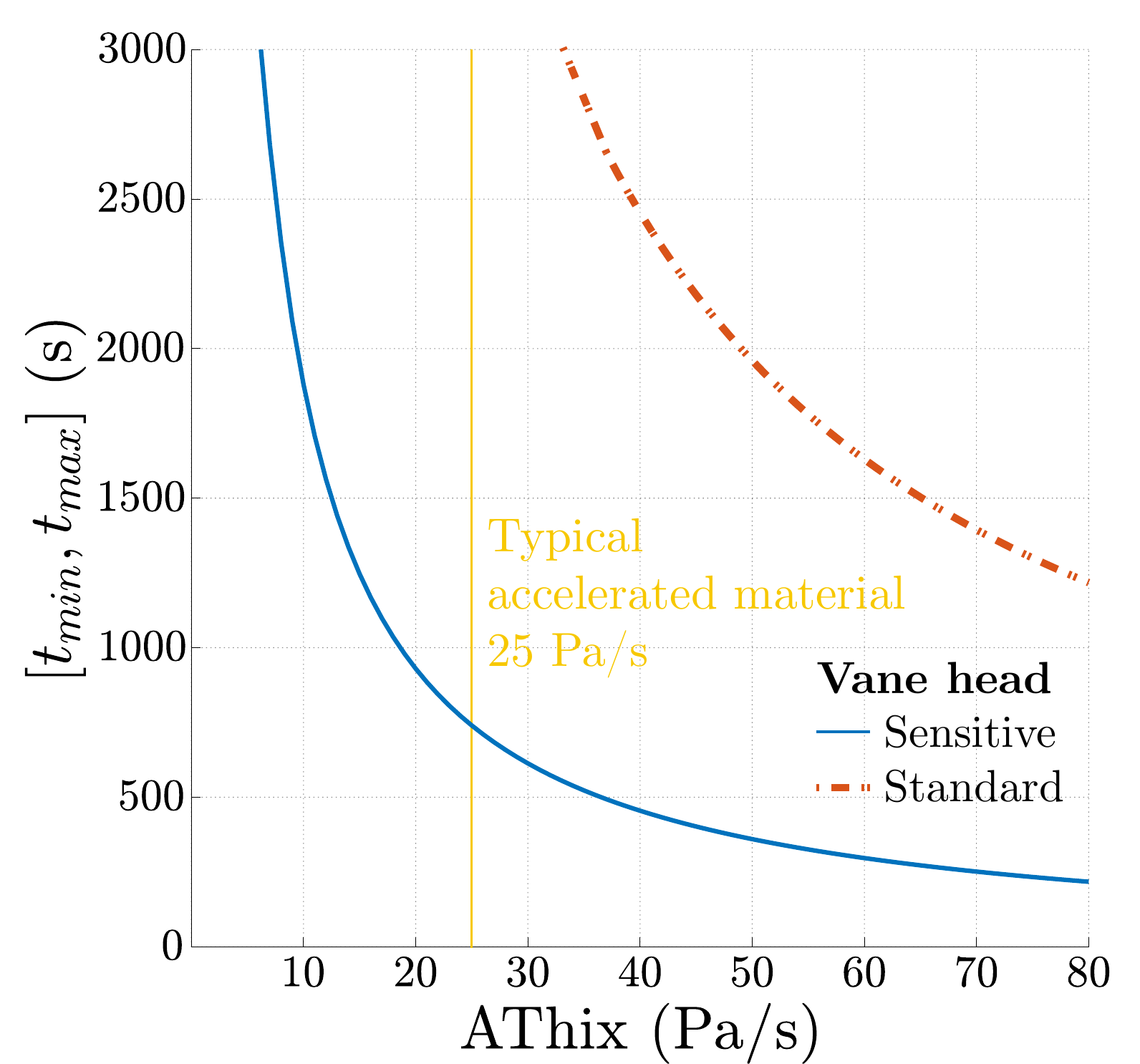} 
    \caption{Measurement window size $[t_{min},t_{max}]$ for Sensitive and Standard vane head as a function of $A_{thix}$ with $\tau_0 = 1000 Pa$.} 
    \label{fig:limite_fenetre_typique_scisso} 
\end{figure} 
Figure \ref{fig:limite_fenetre_typique_scisso} represents the span of the measurement window as a function of $A_ {thix}$ for a material with $\tau_0=1000 Pa$, and this for two measurement vane heads (Sensitive and Standard) indicated by the blue and red curves. The curves are decreasing, because the higher $A_{thix}$, the faster the limit ($\tau_{max},t_{max}$) is reached. 
For $A_{thix}= 25 Pa/s$, the size of the window is 750 s (12.5 min) for the Sensitive vane head and more than 45 min for the Standard vane head which is largely sufficient.\\
The delay between tests $\Delta t_{test}$, as we have seen (§\ref{subsection:test_delay_paillasse}) is either $t_{min}$ the minimum time to carry out a test, or the ratio between the precision of the instrument and the structuration rate $\frac{\delta_{\tau}}{A_{thix}}$. For inline situations, the tests can be sequenced and spaced freely, because the material is produced continuously. It is therefore not $t_{min}$ which imposes the delay, but $\frac{\delta_{\tau}}{A_{thix}}$.\\ 

\begin{figure}[h] 
    \centering 
    \includegraphics[width=0.6\columnwidth]{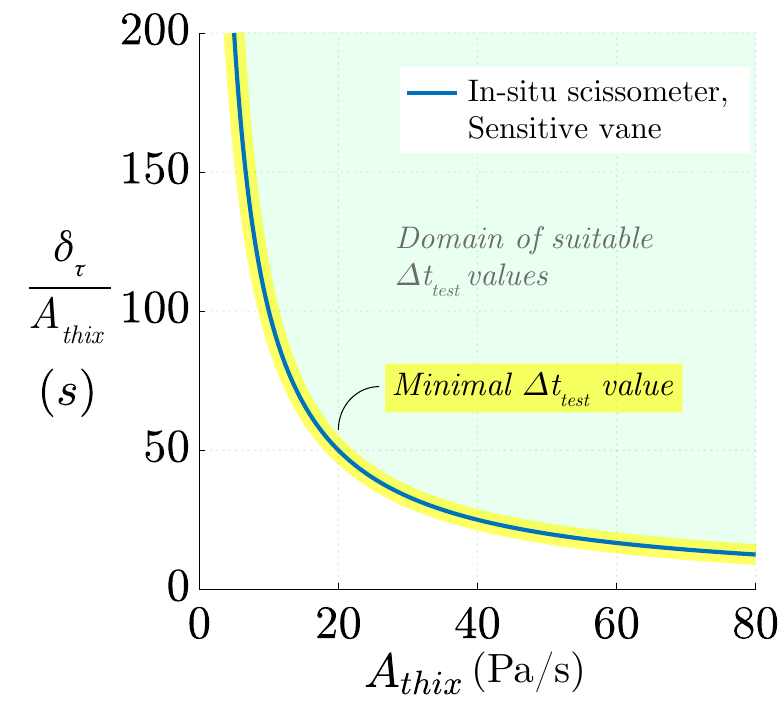} 
    \caption{$\Delta t_{test}$ as a function of $A_{thix}$ for Sensitive vane head.} 
    \label{fig:delta_t_test_athix_calibresensible} 
\end{figure}
This is illustrated for the sensitive vane head (precision $\delta_{\tau}=1kPa$) in Figure \ref{fig:delta_t_test_athix_calibresensible} where the continuous line curve highlighted in yellow represents the minimum value of $\Delta t_{test}$ based on $A_{thix}$.\\

\begin{figure}[h] 
    \centering
    \includegraphics[width=0.6\columnwidth]{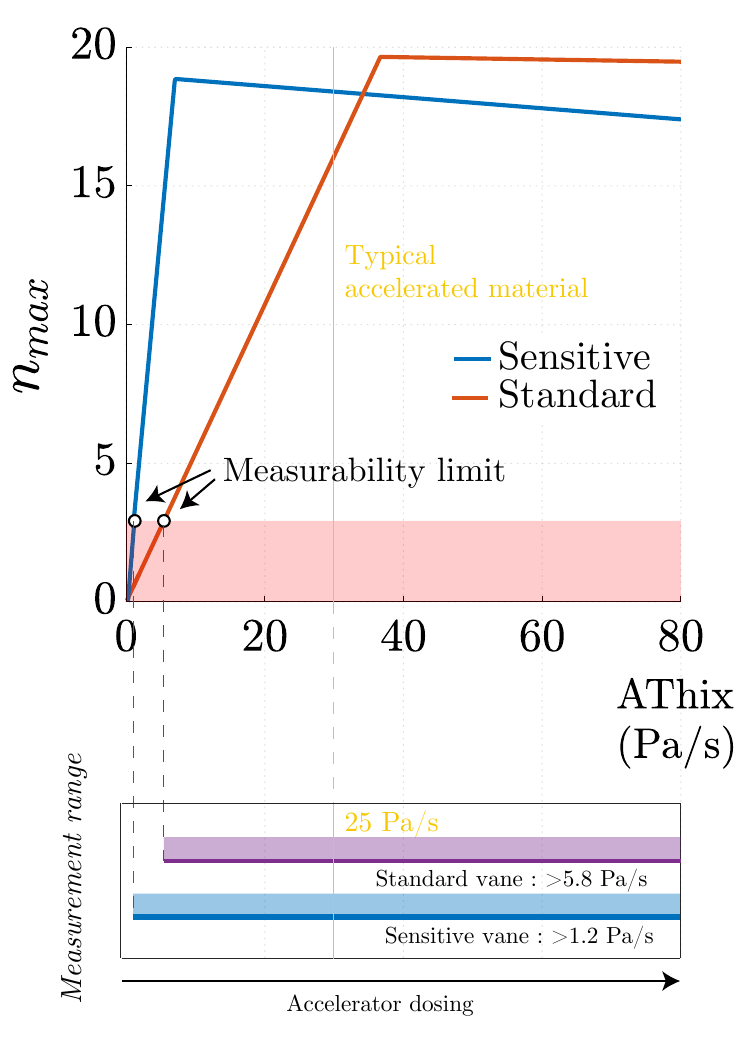} 
    \caption{Number of maximum measurements $n_{max}$ as a function of $A_{thix}$ for the Sensitive and Standard vane heads.} 
    \label{fig:nmax_scisso_sensible_standard} 
\end{figure} 

The evolution of the maximum achievable number of trials $n_{max}$ as a function of $A_{thix}$ with $\tau_0 = 1000 Pa$ is illustrated in Figure \ref{fig:nmax_scisso_sensible_standard}, and follows two successive linear trends. Since $n_{max}$ is inversely proportional to $\Delta t_{test}$, which is inversely proportional to $A_{thix}$, $n_{max}$ is proportional to $A_{thix}$. For the Sensitive vane head, the first part concerns structuring rates below $7.5 Pa/s$. The higher $A_{thix}$, the closer the trials can be to each other. The second part illustrates the limit of the $\tau_{max}$ measurement capacity of the vane head used. 

At the bottom of Figure \ref{fig:nmax_scisso_sensible_standard}, we have represented the possible measurement range of $A_{thix}$ for each vane head, that is say for $n \geq 3$. It can thus be seen that the two vane heads make it possible to theoretically measure very high structuring rates, in the range of hundred Pascals per second. 

\section{Conclusion}
Two-component concrete 3D printing is a very promising process, permitting low power consumption at a high flow rate even for highly viscous mortar, since its yield stress remains low during the pumping. An adapted accelerator permits continuously structuring the mortar, by increasing its yield stress, from a liquid behavior to a sufficiently solid behavior, usually with a larger build-up rate than alternative processes. This allows for more complex geometries.  However, it raises the question of choosing the precise dosage of accelerator that should be high enough to avoid catastrophic failure, but low enough to avoid the so-called "cold joints". An efficient measure of the build-up rate of the mortar is then really needed for efficient quality control. To do so, simple tools allowing the yield stress measurements need to be proposed and assessed, in the same trend that the so-called \textit{Fifty-cent rheometer} concept as defined in \cite{roussel_fifty-cent_nodate}. Another question concerns the replicability of measurement made using benchtop measurement and small mortar specimens to the real material production during printing. This paper tackles those questions 
by introducing a new simple inline metrology for assessing the build-up rate of highly reactive mortars.
For this purpose, a pocket shear vane tester is tested for the first time and shown to meet the requirements and prefigure a simple online test for systematic print qualification. 
We demonstrate that the geometry of the pocket hand vane is more appropriate for our particular problem than classical vane geometry, since mobilizing smaller quantities of mortar, and enables in-situ shear yield stress measurements even on freshly printed layers. The various vane geometries allow for measuring the yield stress from 1kPa to 250 kPa also perfectly meeting the requirements.
Original protocols are proposed and tested, and the main results can be summarized :
\begin{itemize}
    \item The yield stress values measured using a pocket hand vane are repeatable and similar to the values obtained using a classical hand vane.
    \item A linear evolution of the parameter $A_{thix}$ (see Eq.\ref{eq:athix}) is relevant to model the build-up rate, i.e. the shear yield stress evolution over the first hour, for mortars accelerated using an aluminum sulfate accelerator.  
    \item The inline build-up rate is significantly higher, up to 5 times higher than the build-up rate $A_{thix}$ of samples prepared using a benchtop mixer.
    \item The build-up rate increases with higher accelerator dosage, which demonstrates the potential of the 2K strategy in adjusting material properties on the fly.
    \item Using the pocket vane, the shear yield stress can be directly measured on printed laces, or in recipients filled by the extrusion head.
    \item The pocket hand vane enables inline measurements of $A_{thix}$ in the range of hundred Pascal per second. 
\end{itemize}

To conclude the pocket hand vane appears as a good contender for a \textit{Fifty-cent rheometer} for 2K printable micro-mortars ($\phi_{max}=0.8mm$) and will be widely and systematically evaluated in our next printing sessions. It would provide quantitative values of the mechanical strength and their evolution that can be used for simulating the process and ensuring successful printing. Future work should also focus on the applicability of the pocket shear vane to printable concrete including larger particles. 

\section*{Appendix}
In this Appendix, we provide the measurement data on the printer-filled and printed lace measurement comparison (see Figure \ref{fig:scisso_boudin_vs_recipient}).

\begin{table}[h!]
\centering
\begin{tabular}{|l|l|l|}
\hline
Yield stress $\tau_{cr}$ (kPa) & Age $t$ (s) & Vane head \\ \hline
10                         & 320         & Sensitive \\ \hline
11                         & 445         & Sensitive \\ \hline
16                        & 480         & Sensitive \\ \hline
35                        & 543         & Standard \\ \hline
25                        & 570         & Standard \\ \hline
25                       & 580         & Standard \\ \hline
40                        & 694         & Standard \\ \hline
40                        & 706         & Standard \\ \hline
45                        & 730         & Standard \\ \hline
\end{tabular}
\caption{Experimental data of printer-filled container measurements.}
\label{table:appendix1_printerfilled}
\end{table}

\begin{table}[h!]
\centering
\begin{tabular}{|l|l|}
\hline
Yield stress $\tau_{cr}$ (kPa) & Age $t$ (s) \\ \hline
20                        & 549         \\ \hline
30                        & 567         \\ \hline
35                        & 582         \\ \hline
40                       & 651         \\ \hline
35                         & 669         \\ \hline
45                         & 669         \\ \hline
50                         & 769         \\ \hline
40                        & 781         \\ \hline
50                        & 792         \\ \hline
\end{tabular}
\caption{Experimental data of printed lace measurements. The vane head used is Standard.}
\label{table:appendix2_printedlace}
\end{table}

 \section*{Acknowledgment}
This work was made in the framework of Leo Demont's Ph.D.   Leo Demont's thesis is funded by Ecole des Ponts ParisTech and  \href{https://www.buildin-enpc.fr/}{Build'in}, a technological platform of its Co-Innovation Lab. The authors also acknowledge the fruitful collaboration with XTreeE, the technology provider of the extruder, and Jean-Michel Pereira for suggesting the use of a pocket shear vane.

\bibliography{Bibliography_AThix}

\end{document}